\documentclass[12pt,a4paper]{article}
\usepackage[utf8]{inputenc}
\usepackage[T1]{fontenc}

\usepackage{amsmath}
\usepackage{amssymb}
\usepackage{tabularx}
\usepackage{authblk}
\usepackage[title]{appendix}
\usepackage{graphicx}
\usepackage{color} 
\usepackage[linktoc=page]{hyperref}
\hypersetup{
	colorlinks=true,
	linkcolor=blue,
	bookmarks=true,
	citecolor=[rgb]{0.54,0,0}
}
\author{Yoav Zigdon}
\affil{{\normalsize \textit{School of Physics and Astronomy, Tel Aviv University, Ramat Aviv, 69978, Israel}} \\
	{\normalsize yoavzi(at)tauex.tau.ac.il}}
\date{}

\title{Superball of Strings}

\begin{document}
	\maketitle
	\begin{abstract}
	I solve the equations of the low-energy limit of string theory to obtain a solution corresponding to a microcanonical ensemble of highly-excited superstrings. This ``Superball of Strings'' is a static, spherically symmetric ``fuzzball'' of BPS strings with a size set by a random walk scaling.  The solution can be embedded in string theory in a significant part of parameter space. While the solution does not constitute a Lorentzian interpretation for a Euclidean, horizonless solution by Chen, Maldacena, and Witten, a few connections are noted. A singular extremal black hole and the Superball of Strings exist as Supergravity solutions with the same asymptotic boundary conditions; however, I argue that the latter describes generic BPS microstates.  
\end{abstract}
\newpage 
\tableofcontents
\section{Introduction}
String theory admits realistic black hole solutions such as Kerr when suppressing string-loop and higher-curvature corrections. However, such black hole solutions are incompatible with the idea that non-trivial structure resides on the scale of the event horizon, which is supported by the inverse of Hawking's argument \cite{Hawking:1975vcx}, the strong subadditivity inequality \cite{Mathur:2009hf},\cite{Almheiri:2012rt} and an estimation of the expectation value of the occupation number operator of field excitations near the horizon in typical states \cite{Marolf:2013dba}.

Reference \cite{Mathur:2009hf} showed that if a black hole solution with a smooth horizon exists, and if ``niceness conditions'' are met, which include local interactions, small curvature invariants, small energy density and small pressure on Cauchy slices, then the entanglement entropy of Hawking radiation with the remaining black hole increases as a function of evolution step, even when incorporating small corrections to Hawking's computation of the state of the radiation. This result sharpens the tension between the principles of unitary evolution, local interactions, and the equivalence principle.  

String theory also admits solutions corresponding to bound states of strings and branes with no event horizons, which nonetheless share features with black hole solutions, where two examples are a large gravitational redshift for waves propagating from the region far away to the region near the surface of these states, and approximately common geometry far away from the respective objects. For a review posted in 2022, see \cite{Bena:2022rna}. These constructions are qualitatively consistent with the arguments for structure at the scale of the horizon that are cited above.  
More specifically, the radiation is emitted from the surface of the bound states, rather than produced due to quantum fluctuations in the Hartle-Hawking vacuum. In general, the absence of an event horizon implies that there is no obstruction for information to pass from the region near the core of the bound state to the faraway observer. Additionally, unlike the black hole solution, the bound states in question are completely regular.  

String theory may contain bound states of this kind that are 0-BPS, uncharged, and rotating, with compactness similar to that of the Kerr black hole, making them relevant to nature. Potential observational imprints of their existence or absence are ``special'' features in tiny variations in electromagnetic interference patterns, induced by gravitational waves emitted due to the collision of a binary of bound states, where the word ``special'' means that the features do not appear in the gravitational waveforms associated with the collision of two Kerr black holes.  

Another idea related to the current work is that black holes transition into highly-excited fundamental strings when the curvature of their horizons approaches the string scale curvature \cite{Bowick:1985af}, \cite{Horowitz:1996nw}. This idea was corroborated by extrapolations of the entropy (among other thermodynamic quantities) of the respective objects. However, some exceptions that do not require such extrapolations are worldsheet constructions describing low-dimensional target spaces~\cite{Giveon:2005mi}. An implication of this transition about the fate of primordial black holes in the early Universe is that rather than becoming baby Universes~\footnote{Baby Universes bring about information loss for observers in the parent Universe, though they arise in the worldsheet of the string. Arguments against their inclusion in the large-$N$ limit of AdS/CFT have appeared in \cite{Engelhardt},\cite{Kudler-Flam:2025cki}.} or remnants of unbounded degeneracy, as suggested by postmodern approaches to theoretical physics, they have evaporated into radiation because highly-excited strings decay by emitting short strings. Related literature to the black hole/string transition will be reviewed in Section~\ref{sec:com}. 

The objective of this paper is to construct a supersymmetric solution to Supergravity equations, which can be embedded in string theory compactified on a five-dimensional manifold, corresponding to a microcanonical ensemble of highly excited BPS strings. 
A starting point is the set of solutions of fundamental strings coupled to Supergravity, which are parametrized by the embeddings of BPS strings  \cite{Callan:1995hn},\cite{Dabholkar:1995nc}. A version of these solutions with a five-dimensional compact manifold is reviewed in Section~\ref{sec:String}. One can use the conventional quantization of the string and perform an ensemble average of source distributions in the microcanonical ensemble. Similar computations have appeared in the literature in the context of the D1-D5 system  \cite{Alday:2006nd},\cite{Balasubramanian:2008da},\cite{Raju:2018xue}, the NS5-F1 and NS5-P systems \cite{Martinec:2023xvf},\cite{Martinec:2023gte},\cite{Martinec:2024emf}. One of the conclusions of \cite{Raju:2018xue} is that the D1-D5 solution contains a circle that shrinks below the Planck scale in a non-trivial part of target space, which implies that it is not directly possible to embed this particular solution in string theory. Reference \cite{Martinec:2023xvf} showed that a similar phenomenon occurs in the NS5-F1 duality frame; however, the effective field theory of string theory reliably describes the region near the sources in the T-dual NS5-P duality frame. \\
An open question is whether a version of the solution in F1-P can be embedded in string theory. This paper provides a positive answer to this question in a large region in parameter space, for example, in a regime where the asymptotic string coupling is extremely small while holding other parameters fixed. 

One of the advantages of this solution is that it partly fills the dearth of known, spherically symmetric, horizon-free, and singularity-free solutions that exhibit similarities to black holes; the literature about Supergravity and worldsheet solutions of this sort often discusses solutions based on circular profiles of branes in two transverse dimensions, which break spherical symmetry. A partial list of examples of non-spherical solutions is  \cite{Lunin:2001fv},\cite{Bena:2015bea},\cite{Martinec:2017ztd},\cite{Martinec:2025xoy}. Another advantage of a generalized version of solution in subsection~\ref{sec:gen} is that it provides a physical picture for typical states in superselection sectors of the Hilbert space of 1/4-BPS string states with large charges, because such states are ``close to'' the microcanonical ensemble, in the sense that expectation values of simple operators in generic states are approximated by the expectation values in the microcanonical ensemble, up to exponentially small corrections in the entropy.

The calculation leading to this solution is written in Section~\ref{sec:bas}, and a few generalizations to various dimensions are made in subsection~\ref{sec:gen}.  In Section~\ref{sec:com}, several commonalities are pointed out between the Lorentzian solution of the previous section to a Euclidean solution by \cite{Chen:2021dsw}, which is horizon-free, singularity-free, and of approximately the same size. Several local differences are emphasized, too.   
The paper ends with a conclusion and future directions in Section~\ref{conc}. 
	\section{Basics and Calculation}
	\label{sec:bas}
	The spectra of the heterotic string theory and Type II superstring theory compactified on $ S^1_y \times \mathcal{M}$ with $\mathcal{M}=\mathbb{T}^4~\text{or}~\mathcal{M}=K3$ contain  states preserving eight supercharges, with quantized $n_1$  units of winding and $n_{p,1}$ units of momentum with respect to the circle $S^1 _y$. The asymptotic radius of the $S^1 _y$ is denoted by $R_y$, which is taken to be parametrically larger than the string scale: $R_y\gg \sqrt{\alpha'}$. 	The asymptotic string coupling is denoted by $g_s$, and the case where perturbative string theory is applicable asymptotically is considered, namely $g_s \ll 1$. The strings are assumed to be smeared about the $\mathcal{M}$ directions and the volume of the $\mathcal{M}$ is given by
	\begin{equation}
		\text{Vol}(\mathcal{M}) = (2\pi)^4 V_4~.
	\end{equation}
	The case $V_4 \sim (\alpha')^2$ is considered. The asymptotic spacetime where the states live is a product of $S^1 _y \times \mathcal{M}$ times a five-dimensional Minkowski spacetime:
	\begin{align}
	 \mathbb{R}^{1,4} \times S^1 _y \times \mathcal{M}~.
	\end{align}
	 The charges that the BPS states carry are related to the asymptotic moduli and quantized charges as follows
	 \begin{equation}
	 	Q_1 = \frac{g_s ^2 n_1 (\alpha')^3}{V_4}~~,~~ Q_P = \frac{g_s ^2 n_{p,1} (\alpha')^4}{ V_4 R_y^2}~.
	 \end{equation}
	 The total excitation level of the strings $N$ is related to the quantized charges due to the level matching condition, as in
	 \begin{equation}
	 	\label{Levelm}
	 	N = n_1 n_{p,1}~.
	 \end{equation}
	Supergravity equations govern the 1/4-BPS states in regions where the string coupling is weak, the curvature is low in string units, and the sizes of cycles are greater than the string scale. 
	The coordinates $u$ and $v$ are introduced, which are related to time and the dimension along the circle:
	\begin{equation}
		u\equiv\frac{t-y}{\sqrt{2}}~~,~~ v\equiv \frac{t+y}{\sqrt{2}}~~.
	\end{equation}
In the case where the Supergravity approximation is valid, the BPS states are described by a supersymmetric ansatz in which the metric, $B$-field, and the exponential of twice the dilaton are given by \cite{Callan:1995hn},\cite{Dabholkar:1995nc}:
	\begin{align}
		\label{metric}
		\centering
		ds_{10}^2 &= -\frac{2}{Z_1(\vec{x},v)}dv \Big(du+\omega_i(\vec{x},v) dx^i + \frac{\mathcal{F}(\vec{x},v)}{2}dv\Big) + dr^2+r^2 d\Omega_3 ^2 + ds_{\mathcal{M}} ^2~,\nonumber\\
		B^{(2)} &= du\wedge dv -\frac{1}{Z_1(\vec{x},v)} (du+\omega_i(\vec{x},v) dx^i) \wedge dv ~,\nonumber\\
		e^{2\Phi} &= \frac{g_s ^2}{ Z_1(\vec{x},v)}~.
	\end{align}
	The supersymmetric nature of the ansatz implies the existence of a Killing vector $\frac{\partial}{\partial u}$, which is bilinear in the Killing spinor. In subsection~\ref{sec:BH}, an extremal black hole solution carrying winding and momentum charges is reviewed. In subsection~\ref{sec:String}, solutions sourced by oscillating, wound, rotating strings are reviewed. The ``Superball of Strings'' solution is computed in subsection~\ref{sec:ball}, and the verification of the validity of effective field theory about this solution is performed in subsection~\ref{sec:trust}. Subsection~\ref{sec:gen} generalizes the solution in two aspects, among them are solutions associated with different numbers of non-compact dimensions. 
	\subsection{Black Hole Solution}
	\label{sec:BH}
	There exists a supersymmetric, extremal black hole solution carrying $n_1$ units of winding and $n_{p,1}$ units of momentum, which is characterized by
	\begin{equation}
		Z_1 = 1+\frac{Q_1}{r^2}~~,~~
		\omega_i =0~~,~~
		\mathcal{F} = -\frac{2Q_P}{r^2}~~.
	\end{equation}
	Reference \cite{Sen:1995in} considered a closely related, four-dimensional version of this black hole in the low-energy limit of the heterotic string theory compactified on $\mathbb{T}^6$.\footnote{Euclidean rotating solutions whose exponential of minus the on-shell action gives rise to the supersymmetric index of the stringy microstates have appeared in \cite{Chowdhury:2024ngg} and \cite{Chen:2024gmc}.}
		For the black hole solution in the present context, the Ricci scalar reads
	\begin{equation}
		R = -\frac{14 Q_1 ^2}{r^2 (Q_1+r^2)^2}~.
	\end{equation}
	The solution features a horizon and a singularity at the coordinate $r=0$. The near-horizon limit of the six-dimensional metric reads
	\begin{equation}
		\label{NH}
		ds^2 _6 \approx -\frac{2r^2}{Q_1} dv du +\frac{2Q_P}{Q_1}dv^2 +dr^2 + r^2 d\Omega_3^2~. 
	\end{equation}
	The dilaton near the horizon is approximately
	\begin{equation}
		e^{2\Phi} \approx g_s^2 \frac{r^2}{Q_1}~.
	\end{equation}
	Even though the Bekenstein-Hawking entropy of this supersymmetric, extremal black hole solution vanishes, one can associate a Bekenstein-Hawking entropy to a  ``stretched horizon'' - which is defined as a surface where the curvature is of order the string scale	\footnote{Reference \cite{Sen:1995in} introduced a second definition of the concept of the ``stretched horizon'', that the local temperature in the Euclidean continuation of the solution is the Hagedorn temperature of flat spacetime.}, and has the property that the Bekenstein-Hawking entropy of this surface matches the microscopic entropy of the fundamental highly-excited string.
    The Bekenstein-Hawking entropy of a sphere of radius $r$ (times the $S^1 _y \times \mathcal{M}$)  is proportional to the area of that surface in the string frame times $e^{-2\Phi}$ at the surface:
	\begin{align}
		\label{SBH} 
		S_{\text{BH}}(r)=&\frac{e^{-2\Phi}\text{Area}(r) }{4G_N}=\frac{\frac{Q_1}{g_s^2 r^2}\sqrt{\frac{2Q_P}{Q_1}}\times 2\pi^2 r^3  \times 2\pi R_y (2\pi)^4 V_4 }{\frac{1}{4\pi}(2\pi)^7 
			(\alpha')^4} =2\pi\sqrt{2n_1 n_{p,1}}\frac{r}{\sqrt{\alpha'}}~. 
	\end{align} 
	On the other hand, the entropy of a highly-excited Type II string compactified on $\mathcal{M}=\mathbb{T}^4$ is given by
	\begin{equation}
		\label{Smicro}
		S_{\text{micro}}= 2\pi \sqrt{\frac{c}{6} N}=2\pi\sqrt{2n_1 n_{p,1}}~.
	\end{equation}
	The middle expression in Eq.~(\ref{Smicro}) is the Cardy formula for the entropy of states in a CFT of central charge $c$ and a large excitation level $N$, taking the right-movers to be in their ground state and the left-movers highly excited. 
	In the last transition, the central charge was set to $c=8+4=12$ (the superstring has eight bosonic transverse excitations and eight fermionic ones, with the fermions possessing half of the central charge of the bosons). For a $K3$ compact manifold, $c=24$ and $	S_{\text{micro}}= 4\pi\sqrt{n_1 n_{p,1}}$.
	Eqs.~(\ref{SBH}) and (\ref{Smicro}) imply that at a certain radial length scale proportional to the string scale, the Bekenstein-Hawking entropy of the surface matches the microscopic entropy of the fundamental string:
	\begin{equation}
		\label{stretched} 
		r_{\text{stretched}}= \sqrt{\alpha'}~.
	\end{equation}
	To summarize, the two-charge black hole solution has a near-horizon region written in Eq.~(\ref{NH}) and its stretched horizon is written in Eq.~(\ref{stretched}).
	In subsection~\ref{sec:ball}, a much larger solution will be found, which satisfies the same asymptotic boundary conditions, but lacks a singularity and a horizon. A conventional wisdom is that the black hole encapsulates generic high-energy states in any system; this proposition is addressed in the first and final sections of this paper.
	\subsection{Family of String  Solutions}
	\label{sec:String}
	Next, the solutions by Dabholkar, Gauntlett, Harvey, and Waldram (DGHW)  \cite{Dabholkar:1995nc} are reviewed, which are supersymmetric Supergravity solutions with explicit string profile sources carrying $n_1$ winding units and $n_{p,1}$ momentum units.
	The Supergravity fields and string sources are described by the action
	\begin{align}
		I &= \frac{1}{2\kappa_{10} ^2} \int d^{10} x \sqrt{-G} e^{-2\Phi} \Big( R + 4 G^{\mu \nu} \partial_{\mu} \Phi \partial_{\nu} \Phi -\frac{1}{12} H_{\alpha \beta \gamma} H^{\alpha \beta \gamma} \Big)\nonumber\\
		&~ - \frac{\tau_{\text{F1}}}{2} \int d^2 \sigma \Big( \sqrt{-\gamma} \gamma^{ab} G_{\mu \nu} \partial_{a} X^{\mu} \partial_{b} X^{\nu}+\epsilon^{ab} B_{\mu \nu}\partial_a X^{\mu} \partial_{b} X^{\nu}   \Big)~.
	\end{align}
	The tension of the strings is given by
	\begin{equation}
		\tau_{\text{F1}} \equiv \frac{1}{2\pi \alpha'}~,
	\end{equation}
	In Type II superstring theory, the constant $\kappa_{10} ^2$ is given in terms of the string scale to the eighth power:
	\begin{equation}
		\kappa_{10} ^2 = \frac{1}{2} (2\pi)^7 (\alpha')^4~.
	\end{equation}
	Also, define a six-dimensional gravitational coupling parameter 
	\begin{equation}
		\kappa_6 ^2 \equiv \frac{\kappa_{10}^2}{\text{Vol}(\mathcal{M})}~.
	\end{equation}
	The Virasoro constraints are
	\begin{equation}
		G_{\mu \nu} \partial_a X^{\mu} \partial_b X^{\nu} = \frac{1}{2} \gamma^{cd} \gamma_{ab} G_{\alpha \beta} \partial_c X^{\alpha} \partial_d X^{\beta}~.
		\end{equation} 
		The string equation of motion reads
		\begin{equation}
			\nabla^2 _{\text{WS}}X^{\mu} =\Gamma^{\mu} _{\alpha \beta} \gamma^{ab} \partial_a X^{\alpha} \partial_b X^{\beta} -\frac{1}{2} H^{\mu} _{~~\alpha \beta} \epsilon^{ab} \partial_a X^{\alpha} \partial_b X^{\beta}~. 
		\end{equation}
		A gauge choice is that the embedding of the string in the $v$-direction is identical to one of the null coordinates of the worldsheet. Consider a set of multi-strings whose embeddings in target space are denoted by $X_n ^{\mu}$ with $n=1,...,n_1$. 
		One can show that a family of solutions is such that the profiles of the strings in the four spatial directions transverse to the compact manifold are arbitrary functions of $v$:
		\begin{equation}
			X_{jn} = X_{jn} (v)~,~ n=1,...,n_1~,~j=1,...,4~.
		\end{equation} 
		Next, $Z_1,\omega_j$ and $\mathcal{F}$ are tied to the observation vector $\vec{x}$ and the embeddings of the strings $X_n ^i (v)$.
	The Kalb-Ramond $u$-$v$ component equation of motion reads
	\begin{equation}
		\label{Z1}
		\nabla^2 Z_1 = -2\kappa_6 ^2 \tau_{\text{F1}} \sum_{n=1} ^{n_1}\delta^4 (\vec{x}-\vec{X}_n(v))~.
	\end{equation}
	The $B_{uv}$ and $v$-$j$ metric equations (where $j=1,...,4$) imply 
		\begin{equation}
			\label{omega}
		\nabla^2 \omega_j = -2\kappa_6 ^2 \tau_{\text{F1}} \sum_{n=1} ^{n_1} \partial_v X_{nj} \delta^4 (\vec{x}-\vec{X}_n(v))~.
	\end{equation}
	Equations (\ref{Z1})-(\ref{omega}), including one of the Virasoro constraints and the $v$-$v$ metric equation of motion give rise to
\begin{equation}
	\label{F}
	\nabla^2 \mathcal{F} = 2\kappa_6 ^2 \tau_{\text{F1}} \sum_{n=1} ^{n_1} |\partial_v \vec{X}_{n}|^2\delta^4 (\vec{x}-\vec{X}_n(v))~.
\end{equation}
The product that appears in the equations above, $2\kappa_6 ^2 \tau_{\text{F1}}$, can be simplified as
	\begin{equation}
	2\kappa_6 ^2 \tau_{\text{F1}} = (2\pi)^7 \frac{g_s ^2 (\alpha')^4}{(2\pi)^4 V_4} \frac{1}{2\pi \alpha'}=4\pi^2 \frac{g_s^2 (\alpha')^3}{V_4}~. 
\end{equation}
The solutions to Eqs.~(\ref{Z1}),(\ref{omega}) and (\ref{F}) are given by
\begin{align}
	\label{Z1b}
	Z_1 (v,\vec{x}) = 1+\frac{g_s^2 (\alpha')^3}{V_4}\sum_{n=1} ^{n_1} \frac{1}{|\vec{x}-\vec{X}_n (v)|^2}~,
\end{align}
\begin{align}
	\label{omegab}
	\vec{\omega} (v,\vec{x}) = \frac{g_s^2 (\alpha')^3}{V_4}\sum_{n=1} ^{n_1} \frac{\partial_v \vec{X}_n}{|\vec{x}-\vec{X}_n (v)|^2}~,
\end{align}
\begin{align}
	\label{Fb} 
	\mathcal{F} (v,\vec{x}) = -\frac{g_s^2 (\alpha')^3}{V_4}\sum_{n=1} ^{n_1} \frac{|\partial_v \vec{X}_n|^2}{|\vec{x}-\vec{X}_n (v)|^2}~.
\end{align}
 Note that even though expressions (\ref{Z1b}),(\ref{omegab}) and (\ref{Fb}) have singularities at the positions of the strings, the metric, $B$-field and the exponential of the dilaton are completely regular when approaching each string center because they all include $\frac{1}{Z_1}$. \\
Below, the string sources are unified to a single profile consisting of the $n_1$ profile strands, by extending the range of $v$ from $[0,\sqrt{2}\pi R_y]$ to $[0,n_1 \sqrt{2} \pi R_y]$, with $X _j(v)=X_{1j}(v)$ for $v\in [0,\sqrt{2}\pi R_y]$, $X_j (v) = X_{2j} (v)$ for $v\in [\sqrt{2}\pi R_y,2\sqrt{2}\pi R_y]$ until $X_j(v)= X_{jn_1} (v)$ in $v\in [\sqrt{2} \pi R_y (n_1-1),\sqrt{2} \pi R_y n_1]$. The strands are required to satisfy smoothness across $n$, and are periodic with respect to the shift $n\to n+n_1$.
In terms of the unified profile $X_j (v)$, the solutions in Eqs.~(\ref{Z1b})-(\ref{Fb}) are obtained by deleting the sum over $n$ and the subscripts $n$. 

Thus, for different profiles $X_j (v)$, one obtains different excited, wound string solutions. Generalizations of these solutions to other numbers of non-compact dimensions are obtained by replacing the Green's function in four Euclidean dimensions by the Green's function in Euclidean $d$ dimensions. In the next subsection, the solutions above are averaged over in the microcanonical ensemble. 
\subsection{``Superball of Strings'' Solution}
\label{sec:ball}
The goal of this subsection is to obtain a new supergravity solution corresponding to a microcanonical ensemble average of the DGHW string solutions. The microcanonical ensemble is a superselection sector of fixed, large charges $(n_1,n_{p,1})$. The solution in question is computed by averaging the delta-functions appearing on the R.H.S. of Eqs.~(\ref{Z1})-(\ref{F}),  over all BPS states in the superselection sector, each configuration is characterized by its own source profiles. Then, the equations governing the averaged $Z_1,\omega$, and $\mathcal{F}$, Eqs.~(\ref{Z1})-(\ref{F}), are solved with the averaged sources. 

The calculation of the expectation values of these $\delta$-functions,
\begin{equation}
	 \langle \delta^4 (\vec{x}-\vec{X})\rangle  =\langle N | \delta^4 (\vec{x}-\vec{X})  |N\rangle~,
\end{equation}
 is performed in a Hilbert space obtained by quantizing the classical phase space of solutions. This formalism makes averaging convenient; however, one could have performed the average without ever mentioning the quantization of the string (this point is re-emphasized below Eq.~(\ref{computation})).
  The classical moduli space can be parametrized in terms of Fourier amplitudes of the embedding of the string 
 	\begin{align}
 		\label{ModeExpansionClassical}
 		X_j (z,\bar{z}) = x_j -\frac{i}{2}\alpha' p_j \log(|z|^2)&+i \sqrt{\frac{\alpha'}{2}}\sum_{m=1} ^{\infty} \frac{1}{\sqrt{m}} \Big(a_{jm}  \frac{1}{z^m}-a_{jm} ^{*} z^m \Big)\nonumber\\
 		& ~+i \sqrt{\frac{\alpha'}{2}}\sum_{m=1} ^{\infty} \frac{1}{\sqrt{m}} \Big(\tilde{a}_{jm}  \frac{1}{\bar{z}^m}-\tilde{a}_{jm} ^{*} \bar{z}^m\Big)~.
 	\end{align} 
  The center-of-mass-position of the string is $x_j$,  the center-of-mass-momentum of the string is $p_j$, the Fourier amplitudes of left-moving modes are $\{a_{jm},a_{jm} ^{*}\}$, whereas Fourier amplitudes of right-moving modes are $\{\tilde{a}_{jm},\tilde{a} _{jm} ^{*}\}$. Pure left-moving string profiles are characterized by $p_j=0, \tilde{a}_{jm}=0,\tilde{a}_{jm} ^*=0$.   
The coordinates $z,\bar{z}$ on the worldsheet are related to target space coordinates via
\begin{equation}
	z = e^{-i \frac{v\sqrt{2}}{R_y}}~,~ \bar{z} = e^{-i \frac{u \sqrt{2}}{R_y}}~.
\end{equation}
\footnote{More precisely, one chooses a light-cone gauge that ties a null worldsheet coordinate to the target space coordinate $v$, and solves a Virasoro constraint to obtain that the target space coordinate $u$ is linear in the orthogonal worldsheet null coordinate.}  
  One can quantize the transverse excitations of the string by promoting the Fourier coefficients to operators satisfying a suitable algebra. Doing so allows one to write the quantum mode expansion as \cite{Polchinski:1998rq}
\begin{align}
	\label{ModeExpansion}
	\hat{X}_j (z,\bar{z}) = \hat{x}_j -\frac{i}{2}\alpha' \hat{p}_j \log(|z|^2)&+i \sqrt{\frac{\alpha'}{2}}\sum_{m=1} ^{\infty} \frac{1}{\sqrt{m}} \Big(\hat{a}_{jm}  \frac{1}{z^m}-\hat{a}_{jm} ^{\dagger} z^m \Big)\nonumber\\
	& ~+i \sqrt{\frac{\alpha'}{2}}\sum_{m=1} ^{\infty} \frac{1}{\sqrt{m}} \Big(\hat{\tilde{a}}_{jm}  \frac{1}{\bar{z}^m}-\hat{\tilde{a}}_{jm} ^{\dagger} \bar{z}^m\Big)~.
\end{align} 
The notations $\hat{x}_j$, $\hat{p}_j$ are center-of-mass positions and momenta, $\{\hat{a}_{jm},\hat{a}_{jm} ^{\dagger}\}$ annihilate and create left-moving excitations, and $\{\hat{\tilde{a}}_{jm},\hat{\tilde{a}} _{jm} ^{\dagger}\}$ lower and raise right-moving excitations. 
The Hilbert space is a direct sum of Fock spaces, associated to excitations about the vacuum, labeled by the transverse space dimension $j=1,...,4$,  and the mode number $m=1,...,\infty$.
The Hamiltonian reads
\begin{equation}
	\label{L0} 
	\hat{L}_0 = \frac{1}{4}\alpha' \hat{p}^2 + \sum_{m=1} ^{\infty}  \sum_{j=1} ^4 m \hat{a}^{\dagger} _{jm} \hat{a}_{jm}~.
\end{equation}
Below, the degrees of freedom $x_j$ are treated as classical variables and are set to $0$, because a quantum mechanical treatment of them in a microcanonical ensemble spreads the sources across the entire transverse space, since $\hat{x}_j$ are zero modes. For example, if one asks what is the variance of $\hat{X}$ in free string theory, including the zero mode, the result diverges. Similarly, the variables $p_j$ are set to zero classically.  

The expectation value of the delta-function operator in and of itself is ill-defined even in the absence of the zero modes, because it is a composite operator with a UV divergence. Therefore, a renormalization scheme is required to define it. One such renormalization scheme is normal ordering, which is adopted in the following. A discussion about the scheme-independence of the result in a particular limit appears below Eq.~(\ref{rb2}).\\ Now, an identity connecting the delta-function in position space to an integral in momentum space of an exponential is useful:
\begin{align}
	\label{delta}
	\langle : \delta^4 \big(\vec{x}-\vec{X}\big) :\rangle =& \frac{n_1}{(2\pi)^4} \int d^4 k ~\langle :e^{i \vec{k}\cdot (\vec{x}-\vec{X})}:\rangle~.
\end{align}
Before transitioning into an ensemble of fixed excitation level for the strings, one can compute expectation values in the grand-canonical ensemble where the chemical potential $-2\pi i \tau$ is fixed, and then apply an inverse Laplace transform on the result for the expectation value of the normal-ordered delta-function. The expectation value of $:e^{i \vec{k}\cdot \vec{X}}:$ in the grand-canonical ensemble is
\begin{align}
	\label{expikx}
	\langle : e^{i \vec{k} \cdot \vec{X} } :\rangle (\tau)&=   \text{Tr}\Big(e^{2\pi i \tau \hat{L}_0} : e^{i \vec{k} \cdot \vec{X} } :\Big)\nonumber\\
	&=  \text{Tr}\Big( e^{2\pi i \tau \sum_{m=1} ^{\infty} \sum_{j=1} ^{4} m\hat{a}^{\dagger} _{jm} \hat{a}_{jm}} :e^{\sqrt{\frac{\alpha'}{2}} k_j\sum_{m=1} ^{\infty} \frac{\hat{a}_{jm} ^{\dagger} z^m-\hat{a}_{jm} \frac{1}{z^m}}{\sqrt{m}}}: \Big)~.
\end{align}
The last step involved a substitution of Eqs.~(\ref{ModeExpansion}) and (\ref{L0}) in place of $\hat{L}_0$ and $\vec{X}$. The grand-canonical weights have not been normalized with respect to the partition function, because with the choice in Eq.~(\ref{expikx}), one can make the transition to the fixed-charge ensemble more straightforwardly.  
Normal ordering instructs one to place annihilation operators to the right of creation operators:
\begin{align}
	\langle : e^{i \vec{k} \cdot \vec{X} } :\rangle(\tau) &=  \text{Tr}\Big( e^{2\pi i \tau \sum_{m=1} ^{\infty} \sum_{j=1} ^{4} m\hat{a}^{\dagger} _{jm} \hat{a}_{jm}}  e^{\sqrt{\frac{\alpha'}{2}} k_j\sum_{m=1} ^{\infty} \frac{1}{\sqrt{m}} \hat{a}_{jm} ^{\dagger} z^m} e^{-\sqrt{\frac{\alpha'}{2}}k_j \sum_{m=1} ^{\infty}\frac{1}{\sqrt{m}}\hat{a}_{jm} \frac{1}{z^m}}  \Big)~.
\end{align}
Next, one can use a coherent state basis $\{\alpha_{jm}\}$ to evaluate the trace.
\begin{align} 
	\label{computation0} 
	&\langle : e^{i \vec{k} \cdot \vec{X} } :\rangle(\tau)=\nonumber\\
	& \prod_{jm} \int \frac{d^2 \alpha_{jm}}{\pi}\big\langle \alpha_{jm}| e^{2\pi i \tau \sum_{m=1} ^{\infty} \sum_{j=1} ^{4} m\hat{a}^{\dagger} _{jm} \hat{a}_{jm}} e^{\sqrt{\frac{\alpha'}{2}} k_j\sum_{m=1} ^{\infty} \frac{\hat{a}_{jm} ^{\dagger} z^m}{\sqrt{m}}} e^{-\sqrt{\frac{\alpha'}{2}}k_j \sum_{m=1} ^{\infty}\frac{1}{\sqrt{m}}\hat{a}_{jm} \frac{1}{z^m}}|\alpha_{jm} \big\rangle~.
\end{align} 
Some of the properties of coherent states are briefly reviewed:
The relation between a coherent state and occupation number eigenstates is
\begin{equation}
	\label{coh}
	|\alpha_{jm} \rangle = e^{-\frac{1}{2}|\alpha_{jm}|^2} \sum_{n=0} ^{\infty} \frac{\alpha_{jm} ^{n}}{\sqrt{n!}} | n \rangle~.
\end{equation}
The inner product of two coherent states follows from Eq.~(\ref{coh}), and is given by
\begin{equation}
	\langle \alpha_{jm}| \beta_{jm} \rangle = e^{-\frac{1}{2}|\alpha_{jm} - \beta_{jm}|^2}~.
\end{equation}
In particular, the norm of the coherent state in our convention is unity. 
The coherent states are eigenstates of the annihilation operator:
\begin{equation}
	 \hat{a}_{jm} |\alpha_{jm} \rangle = \alpha_{jm} | \alpha_{jm} \rangle~.
\end{equation}
This equation also follows from Eq.~(\ref{coh}).
When acting with an exponential of the number operator on a coherent state, and using Eq.~(\ref{coh}), the result is
\begin{equation}
	e^{ iA \hat{a}_{jm} ^{\dagger} \hat{a}_{jm}} |\alpha_{jm} \rangle = e^{-\frac{1}{2}|\alpha_{jm}|^2} \sum_{n=0} ^{\infty} \frac{(e^{iA}\alpha_{jm}) ^{n}}{\sqrt{n!}} | n \rangle= e^{-\frac{1}{2}\big(1-e^{iA}\big)|\alpha_{jm}|^2}|e^{iA} \alpha_{jm}\rangle~.
\end{equation}
Utilizing the identities above, one can carry on with the computation of the expectation value of $:e^{i \vec{k} \cdot \vec{X}}:$ as follows:
\begin{align} 
	\label{computation} 
		\langle : e^{i \vec{k} \cdot \vec{X} } : \rangle(\tau)&= \nonumber\\
	& =\prod_{j,m} \int \frac{d^2 \alpha_{jm}}{\pi} e^{-\frac{1}{2}(1-e^{2\pi i m \tau}) |\alpha_{jm}|^2  + \sqrt{\frac{\alpha'}{2m}} k_j \big(e^{2\pi i m \tau}\alpha_{jm} ^* z^m - \alpha_{jm} z^{-m}\big) } \nonumber\\
	&  = \frac{1}{\prod_{jm} \Big(1-e^{2\pi i m \tau}\Big)}  e^{-\frac{1}{2}\alpha' k^2 \sum_{m=1} ^{\infty} \frac{e^{2\pi i m \tau}}{m(1-e^{2\pi i m \tau})}}~,
\end{align}
The following integral identity was used in the last step of Eq.~(\ref{computation}):
\begin{equation}
	\int \frac{d^2 \alpha}{\pi} e^{-A |\alpha|^2 + \beta \alpha + \gamma \alpha^*} = \frac{1}{A} e^{\frac{\beta \gamma}{A}}~.
\end{equation}
Eq.~(\ref{computation}) admits an additional interpretation as an average in the classical phase space, parametrized by the amplitudes $\alpha_{jm}$,  with the measure $e^{-\frac{1}{2}(1-e^{2\pi i m \tau}) |\alpha_{jm}|^2}$. For this reason, one need not invoke a quantum field theoretic language to perform the computation. 
Below, the following notation for $2\alpha' \sum_{m=1} ^{\infty} \frac{e^{2\pi i m \tau}}{m(1-e^{2\pi i m \tau})}$ is adopted:
\begin{equation}
	\label{rb2}
	r_b^2 (\tau) = 2\alpha'\sum_{m=1} ^{\infty} \frac{e^{2\pi i m \tau}}{m(1-e^{2\pi i m \tau})}~.
\end{equation}
At this point, the scheme-dependence of the result is discussed. Suppose that instead of normal ordering, one would have introduced a UV cutoff on the worldsheet $\epsilon$. In this case, Eq.~(\ref{rb2}) receives an additional contribution proportional to $-\alpha' \log(\epsilon)$. In a different scheme where one introduces a maximal mode number $m_{\text{max}}$, the additional contribution is proportional to $\alpha' \log(m_{\text{max}})$.  The main point is that in a limit where $\tau \to0$ first, the scaling of $r_b ^2 $ is with $\tau^{-1}$ in Eq.~(\ref{rb2}) washes out the additional logarithmic term. Thus, there is a mathematical limit that eliminates any dependence on the renormalization scheme.\\ 
The quantity $r_b$ will be interpreted as the size of the ``Superball of Strings'' solution.
The grand-canonical expectation value of the normal-ordered exponential is expressible in terms of
\begin{align}
		\label{exp} 
		\langle : e^{i \vec{k} \cdot \vec{X} } :\rangle(\tau)&= Z(\tau) e^{-\frac{1}{4}r_b ^2 (\tau) k^2}~,
	\end{align}
	where
	\begin{equation}
		Z(\tau) = \frac{1}{\prod_{j=1} ^4 \prod_{m=1} ^{\infty}\Big(1-e^{2\pi i m \tau}\Big) }~.
	\end{equation}
	It will shortly turn out that a description of a highly-excited string requires examining $Z(\tau)$ at small $|\tau|$.
	For small $|\tau|$, an approximation of $Z(\tau)$ can be obtained from an identity of the Dedekind-eta function $\eta(-1/\tau)=(-i\tau)^{\frac{1}{2}}\eta(\tau)$. The approximation for $Z(\tau\approx 0)$ is
	\begin{equation}
		\label{Zapprox}
		Z(\tau\approx 0) \approx (-i\tau)^2 e^{-\frac{\pi}{3i \tau}}~.
	\end{equation}
	One can evaluate the entropy of the string at high-excitation level $N$ by applying an inverse Laplace transform on $Z(\tau)$ to extract the density of states:
	\begin{equation}
		\label{rho}
		\rho(N) = -\int_C d\tau e^{-2\pi i N \tau} Z(\tau)~.
	\end{equation}
	The contour $C$ is parallel to the imaginary axis, which is to the right of possible singularities of $Z(\tau)$.  
	 Using Eq.~(\ref{Zapprox}), a saddle point-approximation of the integral in Eq.~(\ref{rho}), which is valid for $N\gg 1$, ties $\tau$ to $N$ as  
	\begin{equation}
		\tau_* \approx \frac{i}{\sqrt{6N}}~,~
	\end{equation}
	and one observes that at large $N$, the following hierarchy applies: $|\tau_*| \ll 1$~. The saddle-point value of the density of states is 
	\begin{equation}
		\rho(N)\approx e^{2\pi \sqrt{\frac{2}{3}N}}~,
	\end{equation}
	and the entropy is its logarithm:
	\begin{equation}
		\label{Entropy} 
		S(N)\approx 2\pi \sqrt{\frac{2}{3}N}=2\pi \sqrt{\frac{2}{3}n_1 n_{p,1}}~.
	\end{equation}
	This matches the Cardy formula with central charge $c=4$ and left-moving excitation level $N$ (with zero right-moving excitation level $N_R=0$.)\\
	To evaluate $\langle : e^{i \vec{k}\cdot \vec{X}} : \rangle(N)$ in a microcanonical ensemble of excitation level $N$, one applies an inverse Laplace transform on the grand-canonical result:
\begin{equation}
	\langle :e^{i\vec{k}\cdot \vec{X}}:\rangle(N) = -\int_{C} d\tau e^{-2\pi i N \tau} 	\langle :e^{i\vec{k}\cdot \vec{X}}:\rangle (\tau)~. 
\end{equation}
The saddle-point approximation works in the present calculation as well; therefore, an approximated expression for $r_b^2 (\tau)$ is 
	\begin{align}
		\label{Size}
		r_b ^2 (\tau_*) \approx -\frac{2\alpha'}{2\pi i \tau_*}\sum_{m=1} ^{\infty} \frac{1}{m^2}=-\frac{\pi \alpha'}{6 i \tau_*}\approx \alpha'\sqrt{\frac{\pi^2}{6}N}~.
	\end{align}
	A recent paper that reached this scaling with $\sqrt{N}$ using similar methods is \cite{Ceplak:2024dxm}.
Using Eqs.~(\ref{Levelm}) and (\ref{Size}), the scale $r_b$ becomes
\begin{equation}
	\label{Sizeb}
	r_b = \sqrt{\alpha'}\Big(\frac{\pi^2}{6} n_1 n_{p,1}\Big) ^{\frac{1}{4}}~.
\end{equation}
The length scale $r_b$ admits a random walk interpretation, with each step of order of $\sqrt{\alpha'}$ and the number of steps $\sqrt{n_1 n_{p,1}}$. Back to the expectation value of the normal-ordered exponential, its saddle-point value is given in
\begin{align}
	\label{exp2} 
	\langle : e^{i \vec{k} \cdot \vec{X} } :\rangle(N)&\approx \rho(N) e^{-\frac{1}{4}r_b ^2 (N) k^2}~.
\end{align}
For the purpose of averaging the Poisson equation of $Z_1$, Eq.~(\ref{Z1}), one can find the expectation value of the delta-function operator in the microcanonical ensemble. Utilizing Eqs.~(\ref{delta}) and (\ref{exp2}), and normalizing the result by dividing by the density of states, one obtains 
\begin{equation}
	\label{Density}
	\langle :\delta^4 (\vec{x}-\vec{X}) :\rangle (N)=  n_1\int \frac{d^4 k}{(2\pi)^4} e^{i \vec{k} \cdot \vec{x}} e^{-\frac{1}{4} r_b^2 (N) k^2} =  \frac{n_1}{\pi^2 r_b^4 (N)} e^{-\frac{r^2}{r_b^2(N)}}~.
\end{equation}
The averaged equation for $Z_1$ reads
\begin{equation}
	\nabla^2 \bar{Z}_1 = -\frac{4Q_1}{r_b^4} e^{-\frac{r^2}{r_b^2}}~. 
\end{equation}
	The solution is given in 
\begin{equation}
	\label{Z1result}
	\bar{Z}_1 = 1+\frac{Q_1}{r^2}\Big(1-e^{-\frac{r^2}{r_b^2}}\Big)~.
\end{equation}
Next, compute
\begin{align}
	\langle : \partial_v X_{i}\delta^4 \big(\vec{x}-\vec{X}\big) :\rangle=& \frac{n_1}{(2\pi)^4} \int d^4 k ~\langle :\partial_v X_i e^{i \vec{k}\cdot (\vec{x}-\vec{X})}:\rangle=0~.
\end{align}
The last equality follows from the parity symmetry of the coherent state integrals. Then, the averaged $\omega_i$ equation (\ref{omega}) implies zero $\bar{\omega}_i$:
\begin{equation}
	\label{omegasol}
	\bar{\omega}_i = 0~.
\end{equation}
Finally, the expectation value of the normal-ordered operator $|\partial_v \vec{X}|^2\delta^4 \big(\vec{x}-\vec{X}\big)$ is
\begin{align}
	\langle : |\partial_v \vec{X}|^2\delta^4 \big(\vec{x}-\vec{X}\big) :\rangle (N)=& \frac{n_1}{(2\pi)^4} \int d^4 k ~\langle : |\partial_v \vec{X}|^2e^{i \vec{k}\cdot (\vec{x}-\vec{X})}:\rangle\nonumber\\
	&=\frac{n_1 \langle :|\partial_v \vec{X}|^2:\rangle}{(2\pi)^4} \int d^4 k ~ e^{i \vec{k}\cdot \vec{x}} e^{- \frac{1}{4}k^2 r_b^2 (N)} = \frac{2n_{p,1}}{\pi^2 r_b^4 (N)} e^{-\frac{|\vec{x}|^2}{r_b^2 (N)}}~. 
\end{align}
In the last step, the following equation was used:
\begin{equation}
	n_{p,1} = \frac{1}{2}\sum_{n=1} ^{n_1} \langle: |\partial_v \vec{X}_{n}|^2 : \rangle~.
\end{equation}
Hence, the averaged equation (\ref{F}) is
\begin{equation}
	\nabla^2 \bar{\mathcal{F}} = -8\frac{Q_P}{r_b^4} e^{-\frac{r^2}{r_b^2}}~,
\end{equation}
and the solution reads
	\begin{equation}
		\label{Fsol} 
		\mathcal{\bar{F}} = -\frac{2Q_P}{r^2}\Big(1-e^{-\frac{r^2}{r_b^2}}\Big)~.
	\end{equation}
		Near the core of the solution, the string and momentum harmonic function approach
	\begin{align}
		\bar{Z}_1 \to 1+\frac{Q_1}{r_b^2}~~,~~
		\bar{\mathcal{F}} \to -\frac{2Q_P}{r_b^2}~~.
	\end{align}
	The main results of this subsection are Eqs.~(\ref{Z1result}),(\ref{omegasol}) and (\ref{Fsol}), which describe the ``Superball of Strings'' solution mathematically. The interpretation of the results is that of a static, spherically symmetric object charged under winding and momentum with respect to the $y$-circle. Moreover, the solution is horizon-free and smooth, with a characteristic size $r_b\gg \sqrt{\alpha'}$. There is more: If the hierarchy of scales $r_b \ll \sqrt{Q_1}\ll \sqrt{Q_P}$ holds, then the geometry near the core is approximately
	\begin{equation}
		ds^2 \approx -\frac{2r_b^2}{Q_1}du dv + \frac{2Q_P}{Q_1} dv^2 + dr^2+r^2 d\Omega_3^2~,
	\end{equation}
	which approximates the near-horizon geometry of the black hole as in Eq.~(\ref{NH}). In particular, probes experience a high gravitational redshift of order $\frac{\sqrt{Q_1}}{r_b}$ when propagating from infinity to the core (for the black hole solution, one replaces $r_b$ by $0$, and the redshift diverges).
	The next subsection examines whether the solution can be embedded in string theory and discusses aspects of the solution within various hierarchies of scales.  
	\subsection{The ``Superball of Strings'' is Trustworthy}
	\label{sec:trust}
	The purpose of this subsection is to verify whether the ``Superball of Strings'' solution is reliable in string theory by checking if (1) the string coupling is everywhere weak, (2) the spacetime curvature is everywhere small in string units, and (3) the $y$-circle is larger than the string scale. The conclusion will be that there exists a regime in parameter space $(g_s,V_4,R_y,n_1,n_{p,1})$ for which the solution is reliable, which is specified below. \\
		Consider the hierarchies  
	\begin{equation}
		g_s\ll 1~,~ \alpha' \ll R_y ^2~,~ V_4 \sim (\alpha')^2~,
	\end{equation}	
	which permit a good effective field theory description in the asymptotics of the spacetime.\\
	(1) The solution is weakly coupled throughout spacetime, with the maximal string coupling being $g_s$ asymptotically. One can see this explicitly by substituting Eq.~(\ref{Z1result}) into the dilaton in Eq.~(\ref{metric})
	\begin{equation}
		e^{2\Phi} =\frac{g_s^2}{\bar{Z}_1}= \frac{g_s^2}{1+\frac{Q_1}{r^2}\Bigg(1-e^{-\frac{r^2}{r_b^2}}\Bigg)}\leq g_s^2\ll1 ~.
	\end{equation}
	(2) The Ricci scalar of the metric (\ref{metric}) reads
	\begin{equation}
		\label{Ricc} 
		R = -\frac{7}{2} \frac{|\nabla Z_1|^2}{Z_1^2} +\frac{2}{Z_1} \nabla^2 Z_1~.
	\end{equation}
	For the ``Superball of Strings'' solution, the Ricci scalar is obtained by plugging Eq.~(\ref{Z1result}) into Eq.~(\ref{Ricc}), which gives 
	\begin{align}
		R = \frac{2A\Big(1-e^{-x^2}(1+x^2)\Big)\Big(-2x^3-A(7+2x)+A(7+x(2+7x))e^{-x^2}\Big)}{r_b^2 x^2 \Big(A+x^2-A e^{-x^2}\Big)^2}~,
	\end{align}
	where
	\begin{equation}
		x\equiv \frac{r}{r_b}~~,~~ A \equiv \frac{Q_1}{r_b^2}~.
	\end{equation}
	 Several limits of $R(A,x)$ are taken below to check whether stringy curvature could occur in the spacetime manifold. 
	When $x\to \infty$ with fixed $A$, 
	\begin{equation}
		R(A,x\to \infty) \approx -\frac{4A  }{r_b^2 x^3}\to 0~.
	\end{equation}
	When $x\to 0$ with fixed $A$, 
	\begin{equation}
		R (A,x\to 0)\approx -\frac{2A x}{(1+A)r_b^2} \to 0~. 
	\end{equation}
	When $x=1$, the function $R(x=1,A)$ receives a maximum at $A\to \infty$ where
	\begin{equation}
		R(x=1,A\to \infty) \approx -\frac{4.1}{r_b^2}~,|R(x=1,A\to \infty)|\ll \frac{1}{\alpha'}~. 
	\end{equation}
	The last parametric inequality assumes $N=n_1 n_{p,1}\gg 1$.
	When $A \to \infty$, the function $R(x,A\to \infty)$ receives a maximum at $x\approx 1.57$, where
	\begin{equation}
		R(x\approx 1.57,A\to \infty) \approx -\frac{5.3}{r_b^2}~,~ |	R(x\approx 1.57,A\to \infty)|\ll \frac{1}{\alpha'}~.
	\end{equation}
	Also, an output of Mathematica for the numerical maximum of $r_b^2 |R(A,x)|$ across $A\geq0, x\geq 0$ is $x\approx 1.57$ and a very large $A$. The conclusion is that the spacetime manifold in the presence of the strings is weakly curved for $n_1 n_{p,1}\gg 1$.\\
	(3) The proper size of the $y$-circle is related to the metric $y$-$y$ component:  
	\begin{equation}
		\label{gyy}
		g_{yy} = \frac{1-\frac{1}{2} \mathcal{F}}{Z_1}~. 
	\end{equation}
	The question is whether the proper size of the $y$ circle is always greater than the string scale, which is equivalent to
	\begin{equation}
		g_{yy} (t,y,\vec{x})\gg \frac{\alpha'}{R_y^2}~~?
	\end{equation}
	For the ``Superball of Stings'' solution, one substitutes Eqs.~(\ref{Z1result}) and (\ref{Fsol}) into Eq.~(\ref{gyy}), which results in
	\begin{equation}
		g_{yy} = \frac{1+ \frac{Q_P}{r^2}\Big(1-e^{-\frac{r^2}{r_b^2}}\Big)}{1+\frac{Q_1}{r^2}\Big(1-e^{-\frac{r^2}{r_b^2}}\Big)}~.
	\end{equation}
	Below, six hierarchies are considered that cover a large part of the parameter space:  
	\begin{enumerate}
		\item  $\sqrt{Q_P}\ll \sqrt{Q_1}\ll r_b$,
		\item $\sqrt{Q_1}\ll \sqrt{Q_P}\ll r_b$
		\item $r_b \ll \sqrt{Q_1}\ll \sqrt{Q_P}$
		\item $\sqrt{Q_1}\ll r_b \ll \sqrt{Q_P}$
		\item $r_b \ll \sqrt{Q_P}\ll \sqrt{Q_1}$
		\item $\sqrt{Q_P}\ll r_b \ll \sqrt{Q_1}$
	\end{enumerate}
	For all hierarchies, when $r\gg \text{max}\{\sqrt{Q_1},\sqrt{Q_P},r_b\}$, it is the case that $g_{yy}\approx 1$ and the proper size of the $y$ circle is $R_y \gg \sqrt{\alpha'}$.\\
	Suppose hierarchies (1) or (2) take place. These arise in particular if the asymptotic string coupling $g_s$ is tuned to be tiny. Whenever $r\ll r_b$, the metric $y$-$y$ component is approximately one:
	\begin{equation}
		g_{yy} (r\ll r_b) \approx 1~. 
	\end{equation}  
	This renders the solution reliable throughout the spacetime because the proper size of the $y$-circle is always approximately $R_y \gg \sqrt{\alpha'}$.\\ For hierarchy (3), when $\sqrt{Q_1}\ll r \ll \sqrt{Q_P}$,
	\begin{equation}
		g_{yy} (\sqrt{Q_1}\ll r \ll \sqrt{Q_P})\approx \frac{Q_P}{r^2}\gg 1~.
	\end{equation} 
	When $r_b \ll r \ll \sqrt{Q_1}$,
	\begin{equation}
		g_{yy}(r_b \ll r \ll \sqrt{Q_1}) \approx \frac{Q_P}{Q_1}\gg 1~.
	\end{equation}
	This continues to be a good approximation when $r\ll r_b$. Therefore, the proper size of the $y$-circle is always parametrically larger than $\sqrt{\alpha'}$. \\ Next, consider the fourth hierarchy. When $r_b \ll r \ll \sqrt{Q_P}$,  
	\begin{equation}
		g_{yy}(r_b \ll r \ll \sqrt{Q_P}) =  \frac{Q_P}{r^2}\gg 1~.
	\end{equation}
	For $\sqrt{Q_1}\ll r \ll r_b$,
		\begin{equation}
		g_{yy}(\sqrt{Q_1}\ll r \ll r_b) =  \frac{Q_P}{r_b^2}\gg 1~.
	\end{equation} 
	Thus again, there is no breakdown of effective field theory for hierarchy (4).
	The hierarchies (5)-(6) pose a danger that the $y$-circle could plunge below the string scale, already far away from the scale $r_b$. For example, in hierarchy (5), when $n_1 \gg n_{p,1}$, 
	\begin{equation}
		g_{yy} (r_d) \approx \frac{\alpha'}{R_y^2}~,~ r_d = \sqrt{\frac{g_s^2 n_1 (\alpha')^4}{R_y ^2 V_4}}~.
	\end{equation}
	The scale $r_d$ above is in the regime $\sqrt{Q_P}\ll r_d \ll \sqrt{Q_1}$ and $r_d\gg r_b$.
	To address this stringy behavior of the $y$-circle, one can apply a T-duality relative to the $y$-circle from the surface where $g_{yy} \sim \frac{\alpha'}{R_y^2}$. A similar procedure was applied in~\cite{Brandenberger:1988aj} in the context of string cosmology.
	The rules of the duality transformations are
	\begin{equation}
		R_y \to R_y '=\frac{\alpha'}{R_y^2} ~,~ g_s \to g_s'=\frac{g_s \sqrt{\alpha'}}{R_y}~, n_1 \leftrightarrow n_{p,1}
	\end{equation}
	\begin{equation}
		Z_1 \to Z_1'=1- \frac{\mathcal{F}}{2}~,~ \mathcal{F} \to \mathcal{F}' = 2(1-Z_1)~.
	\end{equation}
	These rules imply that $Q_1 \leftrightarrow Q_P$ and therefore, take hierarchy (5) to (3), and send hierarchy (6) to (4). The string coupling remains weak throughout in this T-dual background, because
	\begin{equation}
		e^{2\Phi'} = \frac{(g_s')^2}{1-\frac{\mathcal{\bar{F}}}{2}} \leq (g_s')^2~.
	\end{equation}
	Further, the maximal curvature in the geometry remains weak just as in the original frame: $R_{\text{max}}\sim \frac{O(1)}{r_b^2}\ll \frac{1}{\alpha'}$. The proper size of the $y$-circle in the T-dual frame is larger than the string scale when it goes below the string scale in the original frame. The conclusion is the existence of a ``Superball of Strings'' solution in which all validity conditions of effective field theory are met, which implies that in all six hierarchies, the solution discussed in the paper, or the solution attached to a T-dual version thereof (when the hierarchies 5-6 are considered, if a sub-stringy behavior occurs in the original frame), can be embedded in string theory. Moreover, in all cases where the solution is trustworthy, it exists with the same boundary conditions for which the singular black hole solution exists. 
	\subsection{Generalizations}
	\label{sec:gen}
	This subsection generalizes the ``Superball of Strings'' solution in two ways. First, solutions in which the strings are excited in the compact manifold dimensions are obtained, with a size smaller than the one in subsection~\ref{sec:ball}. Second, different numbers of non-compact dimensions $d$ are considered, with spacetime asymptotics $\mathbb{R}^{1,d}\times S^1 _y \times \mathcal{M}_{8-d}$, and different solutions are produced depending on the dimensionality $d$.
	
	The first setup is 1/4-BPS strings carrying $n_{p,1}$ momentum units with the following partition
	\begin{equation}
		n_{p,1} = n_{p,1} ^{\perp} + n_{p,1} ^{\parallel}~,
	\end{equation}  
	where $n_{p,1} ^{\perp}$ denotes the quantized momentum number associated with the transverse directions, and $n_{p,1} ^{\parallel}$ is the quantized momentum number for the compact dimensions $\mathcal{M}$. In other words, the stress-energy tensor on the worldsheet has contributions from both embeddings of the strings in transverse space, and in $\mathcal{M}$, with their integrals providing the quantized momenta numbers. It is assumed that both $n_{p,1} ^{\perp},n_{p,1} ^{\parallel}\gg1$. In the presence of internal excitations of the strings in $\mathcal{M}$, a nonzero stress-energy tensor in target space is induced with legs on $\mathcal{M}$. However, one can smear this stress-energy tensor along the directions of $\mathcal{M}$, and in the absence of winding and momentum modes about these directions, the smeared components of the target-space stress-energy tensor along $\mathcal{M}$ vanish, justifying using the same Supergravity ansatz in Eq.~(\ref{metric}).
	The solution corresponding to the microcanonical ensemble with excitation level $N= n_1 n_{p,1} ^{\perp}$ remains almost the same. The nonzero averaged harmonic functions read 
	\begin{align}
		\label{Ensemble}
			\bar{Z}_1 = 1+\frac{Q_1}{r^2}\Bigg(1-e^{-\frac{r^2}{r_{b,\perp}^2}}\Bigg)~~,~~
			\mathcal{\bar{F}} = -\frac{2Q_P}{r^2}\Bigg(1-e^{-\frac{r^2}{r_{b,\perp}^2}}\Bigg)~~,
	\end{align}    
	where the size is given by
	\begin{equation}
		\label{Sizec}
		r_{b,\perp} = \sqrt{\alpha'}\Big(\frac{\pi^2}{6} n_1 n_{p,1}^{\perp}\Big) ^{\frac{1}{4}}~.
	\end{equation}
	Since $n_{p,1} ^{\perp}<n_{p,1}$, the size of this solution is smaller than the one in subsection~\ref{sec:ball}.\\
	The second mode of generalization considers $d$ non-compact dimensions, including time. One can generalize the solution by solving the appropriate sourced ``Poisson'' equations with boundary conditions of suitable falloffs far away and regularity at the center. The result is
	\begin{align}
		\bar{Z}_1 = 1+\frac{Q_1}{r^{d-3}} f_d\Big(\frac{r}{r_{b,\perp}(d)}\Big) ~,~ \bar{\mathcal{F}}=-\frac{2Q_P}{r^{d-3}} f_d\Big(\frac{r}{r_{b,\perp}(d)}\Big)~,
	\end{align}
	where
	\begin{align}
		f_d (x) \equiv 1-\frac{\Gamma\Big(\frac{d-1}{2},x^2\Big)}{\Gamma\Big(\frac{d-1}{2}\Big)}+\frac{x^{d-3}e^{-x^2}}{ \Gamma\Big(\frac{d-1}{2}\Big) }~.
	\end{align}
	The incomplete Gamma function, which appears in the definition of $f_d(x)$, is defined in
	\begin{equation}
		\Gamma(a,x) \equiv \int_x ^{\infty} e^{- t} t^{a-1} dt~.
	\end{equation}
	Also, the size is 
		\begin{equation}
		\label{Sized}
		r_{b,\perp}(d) = \sqrt{\alpha'}\Big(\frac{2\pi^2}{3(d-1)} n_1 n_{p,1}^{\perp}\Big) ^{\frac{1}{4}}~.
	\end{equation}
	In the special case of $d=4$ non-compact dimension, the solution reads
	\begin{equation}
		\bar{Z}_1 = 1+ \frac{Q_1}{r} \text{Erf}\Big(\frac{r}{r_{b,\perp}(4)}\Big)~,~\mathcal{\bar{F}} = -\frac{2Q_P}{r}\text{Erf}\Big(\frac{r}{r_{b,\perp}(4)}\Big)~.
	\end{equation}
	It would be interesting to produce a non-BPS Lorentzian solution sourced by strings in $3+1$ non-compact dimensions corresponding to the microcanonical ensemble;  the error-function above might play a role in such a solution, replacing the Schwarzschild blackening factor $1-\frac{2M}{r}$ by $1-\frac{2M}{r}\text{Erf}\Big(\frac{r}{2M}\Big) $.
\section{Comparison with the Chen-Maldacena-Witten Solution}
\label{sec:com}
\subsection{Review of the Horowitz-Polchinski Solution}
Reference \cite{HP} found a solution to the equations of motion derived from the Euclidean Horowitz-Polchinski (HP) effective action. In Type II superstring theory, this effective action reads
\begin{align}
	I_{HP} &= \frac{\beta}{16\pi G_N}\int d^d x \sqrt{G_d} e^{-2\Phi_d} \Big(\partial_{\mu} \chi \partial^{\mu} \chi^* + m^2 \chi \chi^* + \frac{4}{\alpha'}\varphi \chi \chi^*\Big)\nonumber\\
	&~-\frac{\beta}{16\pi G_N}\int d^d x \sqrt{G_d} e^{-2\Phi_d} \Big(R_d -\partial_{\mu} \varphi \partial^{\mu} \varphi+ 4\partial_{\mu} \Phi_d \partial^{\mu} \Phi_d \Big)~.
\end{align}
The notations in $I_{\text{HP}}$ are a mode of the metric time-time component $G_{\tau \tau}$, namely the radion $\varphi=\frac{1}{2} \Big(G_{\tau \tau}-1\Big)$ when $G_{\tau \tau}\approx 1$, and winding string modes that wind once around the Euclidean time circle ``$\chi,\chi^*$''. $G_d$ is the determinant of the metric in the $d$ dimensions transverse to the thermal circle, and $\Phi_d = \Phi - \frac{\varphi}{2}$.
The mass of the winding condensate in Type II superstring theory is given by
\begin{equation}
	m = \frac{\sqrt{\beta^2-\beta_H^2}}{2\pi \alpha'}~,~ \beta_H \equiv 2\pi \sqrt{2\alpha'}~.
\end{equation}
The Euclidean solution is characterized by a normalizable winding condensate, it is spherically symmetric, uncharged, and has entropy that scales with $1/G_N$ at small string coupling. The functions $\chi(r),\varphi(r)$ can be found numerically, and resemble Gaussians near the center; schematically, write 
\begin{equation}
	\text{HP}: ~ \chi=\chi(r)~,~ \varphi=\varphi(r)~.
\end{equation}
The solution is valid in the regime $O(1) g_s ^{\frac{4}{6-d}}\ll \frac{\beta-\beta_H}{\sqrt{\alpha'}}\ll 1$, with the left hierarchy coming from the requirement that the classical action of the solution is large, and the right hierarchy follows from the lightness of the winding modes.
The size of the solution scales with the inverse of the (asymptotic) mass of the winding condensate $m$:
\begin{equation}
	\ell \sim \frac{1}{m} \sim \frac{(\alpha')^{\frac{3}{4}}}{\sqrt{\beta-\beta_H}} ~.
\end{equation}
 In Type II superstring theory with $d=4$ non-compact dimensions, the mass of the solution is approximately
\begin{equation}
	M_{\text{HP}}(d=4,II) \approx 0.05 \frac{\alpha'}{G_N}~.
\end{equation}
The entropy of the Type II solution contains a term that corresponds to a Hagedorn behavior
\begin{equation}
	\label{entropy}
	S\supset  \frac{\beta_H}{4\pi \alpha' G_N}\int d^d x |\chi|^2 = \beta_H M~.
\end{equation}
The HP solution is thought to describe Lorentzian highly-excited, self-gravitating strings near the Hagedorn temperature, as evidenced by a zero coupling computation that shows an agreement between the partition function of free, high-temperature strings in Lorentzian signature, and the Euclidean partition function of free winding modes \cite{Brandenberger:1988aj}. However, there is no known, explicit, Lorentzian solution or a class of solutions to Supergravity in the presence of string sources, which shares the same thermodynamic quantities as the HP solution. 
\subsection{Review of the Chen-Maldacena-Witten Solution}  
In 2021, Chen, Maldacena, and Witten (CMW)~\cite{Chen:2021dsw} applied an $O(2,2)$ transformation on the HP solution in the presence of an additional, spatial circle $S^1 _y$, to generate a solution with winding and momentum charges around that circle. \footnote{A paper that discussed phases of string stars in the presence of such a spatial circle is \cite{Chu:2025fko}.} The winding and momentum charges along $S^1 _y$ are reflected in off-diagonal $B$-field and metric components. The set of $O(2,2)$ transformations considered in \cite{Chen:2021dsw} was required to preserve the winding property around the thermal circle, such that the winding condensate $\chi$ remains the same after the transformation has been completed.   At tree-level in the string coupling, the authors demonstrated that the partition functions of the HP and CMW solutions are identical, because a boundary term determines them. A consequence is the equality of the entropies, but the energy, asymptotic radius of thermal circle $S^1 _{\beta}$, and asymptotic string coupling are modified in a way computed in \cite{Chen:2021dsw}.  

In terms of two of the parameters of the $O(2,2)$ transformation, $\{\alpha,\gamma\} $, the asymptotic inverse temperature of the new solution $\beta$ is given by
\begin{equation}
	\beta = \tilde{\beta} \cosh(\alpha) \cosh(\gamma)~,
\end{equation}
where $\tilde{\beta}$ is the asymptotic inverse temperature of the HP solution. When $\alpha,\gamma\gg 1$, the asymptotic temperature of the charged solution is parametrically smaller than the Hagedorn temperature. 
The same relation ties the asymptotic string couplings squared of the two solutions:
\begin{equation}
	g_s^2 = \tilde{g}_s ^2 \cosh(\alpha) \cosh(\gamma)~.
\end{equation}
Formulas for the Type II solution are given in terms of the data of the HP solution $\{\varphi(r),\chi(r)\}$. The line element, $B$-field, and the exponential of twice the dilaton are: 
\begin{align}
	\label{CMW}
	ds^2 =&  \frac{1+2\varphi}{(1-2\varphi \sinh^2 (\alpha))(1-2\varphi \cosh^2 (\alpha))}d\tau^2\nonumber\\
	&+R_y^2 \frac{1-2\varphi \sinh^2 (\alpha)}{1-2\varphi \sinh^2 (\gamma)}\Big[dy-i\frac{\tanh(\alpha)}{R_y}\frac{1+2\varphi}{1-2\varphi \sinh^2 (\alpha)}d\tau\Big]^2+d\vec{x}^2~,\nonumber\\
	B^{(2)}&= -i\tanh(\gamma ) R_y \frac{1+2\varphi}{1-2\varphi \sinh^2(\alpha)}dy\wedge d\tau~,\nonumber\\
	e^{2\Phi} &= g_s ^2 \frac{1+\varphi}{(1-2\varphi \sinh^2 (\alpha))(1-2\varphi \sinh^2 (\gamma))}~.
\end{align}
Below, an interesting aspect is a heterotic version of this solution, which is reviewed in a particular BPS limit shortly.  Define
\begin{equation}
	Q_L \equiv \frac{n_{p,1}}{R_y}+\frac{n_1 R_y}{\alpha'} ~,~ Q_R \equiv \frac{n_{p,1}}{R_y}-\frac{ n_1R_y}{\alpha'}~.
\end{equation} 
Eq.~(5.36) of \cite{Chen:2021dsw} ties the entropy of the charged solution to its energy $M$ and $Q_L,Q_R$ in the following manner:
\begin{equation}
	\label{S} 
	S = 2\pi \sqrt{\alpha'}\Big(\sqrt{M^2-Q_L^2}+\frac{1}{\sqrt{2}} \sqrt{M^2-Q_R^2}\Big)~.
\end{equation}
A BPS and an extremal limit of the solution is $M \to Q_R$.
In this limit, the entropy formula (\ref{S}) becomes
\begin{equation}
	 \label{BPS}
	 S \to 4\pi \sqrt{|n_1 n_{p,1}|}~,
\end{equation}
which is the entropy of an ensemble of a highly-excited heterotic string with a $S^1 _y \times K3$ compact manifold.
\subsection{Comparison}
Since the conventional HP solution is expected to match an ensemble of highly-excited, self-gravitating Lorentzian strings, one could anticipate that its charged version corresponds to a Lorentzian ensemble average of charged string solutions. In this subsection, several common features and differences are outlined between the Lorentzian solutions in subsections~\ref{sec:ball},\ref{sec:gen}, and the CMW solution, leading to the conclusion that the ``Superball of Strings'' is not a Lorentzian continuation of the CMW solution.\\
The R.H.S. of Eq.~(\ref{BPS}) is larger than the entropy of the microcanonical ensemble computed in Eq.~(\ref{Entropy}) by a factor of $\sqrt{6}\approx 2.45$. The latter microcanonical entropy is the entropy associated with the Lorentzian solution of section~\ref{sec:ball}. One can achieve a perfect agreement by working with the heterotic string theory compactified on $K3$, where the central charge of the worldsheet theory is $24$, exciting momentum modes about the $K3$ manifold.  This type of generalized solution was discussed in subsection~\ref{sec:gen}. \\
When approaching extremality, the charged solution is expected to make a transition to a free-string phase, whose size admits a random walk scaling:
\begin{equation}
	\label{randomWalk}
	\ell_{\text{free}} \sim \sqrt{\alpha'} (n_1 n_{p,1}) ^{\frac{1}{4}}~.
\end{equation}
This agrees with the scaling of the size of the Lorentzian solution of subsection~\ref{sec:ball} Eq.~(\ref{Sizeb}). However, off-extremality, the Euclidean CMW solution is smaller than the R.H.S. of Eq.~(\ref{randomWalk}) due to self-gravity:
\begin{equation}
	\ell_{\text{CMW}} < \sqrt{\alpha'} (n_1 n_{p,1}) ^{\frac{1}{4}}~.
\end{equation}
Although it does not make sense to directly compare the non-BPS CMW solution to the Lorentzian BPS solution of subsection~\ref{sec:ball}, a comment is that considering the generalization of the solution in subsection~\ref{sec:gen}, where the size has a scaling
 \begin{equation}
 	r_{b,\perp} \sim \sqrt{\alpha'} (n_1 n_{p,1} ^{\perp}) ^{\frac{1}{4}}~,
 \end{equation}   
and one can tune $n_{p,1} ^{\perp}$ to obtain a close approximation to $\ell_{\text{CMW}}$.\\ Additionally, when $\text{max}\{Q_1,Q_P\} \ll r_b ^2$, the redshift factor in the Lorentzian solution is parametrically smaller than one, similar to the CMW solution. Other common features are the spherical symmetry and the flatness of the transverse space.

However, a difference between the CMW solution and the solution of subsection~\ref{sec:ball} is that an analytical continuation of the metric, $B$-field, and dilaton of the latter, with $t\to i\tau$, does not bring about Eqs.~(\ref{CMW}). Moreover, the field $\chi$, which through Eq.~(\ref{entropy}), is akin to the entropy density of the strings in CMW's solution, is different from the density of strings of the Lorentzian solution encoded in the observable $\langle:\delta^4 (\vec{x}-\vec{X}) :\rangle$ in Eq.~(\ref{Density}). For example, the asymptotic falloff of $\chi$ is exponential in the radial coordinate, $\chi(r\to \infty)\to e^{-m r}$, whereas the density of strings of the Lorentzian solution is Gaussian in $r$, as in $e^{-\frac{r^2}{r_b^2}}$. Reference \cite{HP} Eq.~(2.13) showed that a canonical ensemble expectation value of a quantity akin to the density of strings is a conventional exponential of the required form. However, near the center of space, the HP or CMW function $\chi(r)$ deviates from a pure exponential, and the aforementioned relation between $\chi$ and the density of strings explained in \cite{HP} does not account for this deviation. It would be interesting to pinpoint a specific ensemble whose entropy density of strings agrees with the entropy density associated with the profile $\chi=\chi(r)$ of the Euclidean winding condensate of the HP or CMW solution. Such a different ensemble would solve the problem of determining the Lorentzian continuation of the CMW solution. \\ To recapitulate, several differences were pointed out between the CMW solution and the ``Superball of Strings'' solution, even though they share the same charges, with comparable sizes and entropies. 
\section{Conclusion}
\label{conc} 
In this paper, a reliable solution to string theory is presented, whose physical picture is that of a ball of BPS strings with a root-mean-square value given by a random-walk formula. The solution was obtained by performing an ensemble average in a superselection sector of highly excited strings that preserve eight supercharges. Since generic BPS microstates with large charges are almost indistinguishable from the microcanonical ensemble, such a solution provides a weakly coupled, weakly curved description of generic stringy microstates. Several future directions are written. First, generalizing the solution to less supersymmetric systems is important for approaching a phenomenologically relevant model of stringy bound states. The ``Superball of Strings'' may occur in a Universe that admits a spatial compact dimension, as in the scenario of~\cite{Montero:2022prj}; it would be interesting to explore its phenomenological implications. In addition, rotating versions of the solution can be constructed in principle using the methods of~\cite{Martinec:2023xvf}. Furthermore, a specific analytical continuation of the rotating solution could potentially allow one to evaluate the supersymmetric index of the stringy microstates in the spirit of~\cite{Chowdhury:2024ngg}, \cite{Chen:2024gmc}, albeit utilizing a horizon-free solution. Finally, finding a particular ensemble that provides a Lorentzian continuation of the CMW solution is an interesting prospect. 
\subsection*{Acknowledgements}
I thank Ofer Aharony, Sunny Itzhaki, David Kutasov, Amit Sever, Nadav Shrayer, and Paul Steinhardt for discussions. Special thanks to Emil Martinec for useful discussions. I also thank Don Marolf for posing a question about the reliability of the F1-P solution, which is answered in this work. YZ is supported by the Israel Science Foundation, grant number 1099/24.

\end{document}